\documentclass[journal]{IEEEtran}

\usepackage[english]{babel}
\usepackage[utf8]{inputenc}
\usepackage{amsmath}
\usepackage{amsthm}
\usepackage{amsfonts}
\usepackage{mathtools}
\usepackage{graphicx}
\usepackage{float}
\usepackage[justification=centering]{caption}
\usepackage{hyperref}
\usepackage{dsfont}
\usepackage{bbm}
\usepackage{multirow}
\usepackage{flushend}
\usepackage[font=small,labelfont=bf,justification=justified]{caption}

\DeclareMathOperator*{\tr}{\text{tr}}

\title{\LARGE \bf
Tube-based Guaranteed Cost Model Predictive Control Applied to Autonomous Driving Up to the Limits of Handling
}

\author{Carlos M. Massera$^{1,3}$, Tiago C. dos Santos$^{3}$, Marco H. Terra$^{2}$ and Denis F. Wolf$^{3}$
\thanks{$^{1}$Lyft Inc., 2300 26th St, San Francisco, CA, USA 94107 {\tt\small cmassera@lyft.com}}%
\thanks{$^{2}$São Carlos School of Engineering, University of São Paulo, Avenida Trabalhador São-carlense, 400, São Carlos, Brazil {\tt\small terra@sc.usp.br}}%
\thanks{$^{3}$Institute of Mathematics and Computer Science, University of São Paulo, Avenida Trabalhador São-carlense, 400, São Carlos, Brazil {\tt\small massera@icmc.usp.br, tiagocs@icmc.usp.br, denis@icmc.usp.br}}%
}

\begin{document}
\maketitle
\thispagestyle{empty}
\pagestyle{empty}

\begin{abstract}

The development of control techniques to maintain vehicle stability under possible loss-of-control scenarios is essential to the safe deployment of autonomous ground vehicles in public scenarios.
In this paper, we propose a tube-based guaranteed cost model predictive controller for autonomous vehicles able to avoid front and rear tire saturation and to track a provided reference trajectory up to the limits of handling of the vehicle.
Such an approach ensures the vehicle will remain within its safe operational envelope; therefore, guaranteeing both stability and performance of the vehicle, including highly dynamic maneuvers that may be necessary for emergency conditions.
We also propose a new conservative approximation of the nonlinear vehicle dynamics to a linear system subject to norm-bounded multiplicative uncertainties and a  new maximal robust controllable invariant set for vehicle dynamics. It consists of a larger feasible state space region when compared to previously proposed invariant sets.
Finally, we present both simulation and in-vehicle results of the performance of the proposed approach.
\end{abstract}


\section{INTRODUCTION}

Road traffic crashes are the leading cause of death among young people between 10 and 24 years old \cite{world2007youth}. 
Most of these accidents occur when the driver is unable to maintain the vehicle control due to fatigue or external factors resulting in loss-of-control scenarios \cite{national2008national}. 
In recent years, both academia and industry have devoted efforts to the development of safety systems in order to decrease the number of road accidents. 

The development of control techniques to maintain vehicle stability under possible loss-of-control scenarios is  important to current driver assistance systems (DAS) and autonomous ground vehicles (AGV). New technologies are continuously being introduced in commercial vehicles to provide more information and control capabilities for the handling limiting scenarios. One of these technologies is steer-by-wire system \cite{ulrich2013top}, which enables measurements of self-aligning torque and tire-road friction coefficients with higher precision \cite{hsu2010estimation}.

Model predictive control (MPC) is a class of optimization-based control algorithms that use an explicit model of the controlled system to predict its future states \cite{badgwell2015model}. 
Several different fields have applications which use this technique, such as refineries, food processing plants, mining, aerospace, and automotive control \cite{qin2003survey}. 
An MPC minimizes a cost function while maintaining the system states and control inputs within a predefined domain.

Controllable invariant (CI) sets and robust controllable invariant (RCI) sets \cite{blanchini1999set} have been useful in modern control system designs whose interest has increased in the past two decades.
The CI class of sets describes domains of the state space of a dynamic system where there is at least one admissible control input. It maintains the future state space within the same domain. Meanwhile, RCI sets are CI sets where there is uncertainty or disturbances present in the dynamic system.
MPC approaches have used CI and RCI sets to guarantee its infinite horizon stability and feasibility \cite{bemporad1999robust}. Such guarantees are fundamental in the deployment of control systems for safety-critical applications, such as AGVs.

In \cite{beal2013mpc}, the authors  proposed an MPC-based DAS for steer-by-wire vehicles able to ensure vehicle handling limits in coordination with a human driver. They developed a novel  CI set based on the vehicle's maximum steady-state yaw rate and the maximum allowed rear slip angles. This work was later extended to autonomous vehicle control and obstacle avoidance in \cite{funke2017collision}.
In \cite{carvalho2013predictive}, it was designed an MPC for evasive maneuvers based on the iterative linearization of a nonlinear Ackerman model with convexified constraints.
In \cite{falcone2007predictive}, it was developed both nonlinear  and linear time-varying MPC to the reference tracking control problem of autonomous vehicles.
In \cite{raffo2009predictive}, it was formulated a nested controller for reference tracking which consists of two MPC controller: a longer horizon outer-loop based on the kinematic vehicle model to reduce complexity and computational requirements; and an inner-loop which uses the dynamic vehicle model to provide higher fidelity control.
However, such approach do not provide robustness to deal with tire parameter uncertainty.

Tire characteristics vary significantly with temperature \cite{tonuk2001prediction}, wear \cite{braghin2006tyre} and their manufacturing process. Furthermore, they directly impact the vehicle handling limits and the overall system behavior in a multiplicative manner \cite{beal2013mpc} and \cite{rawlings1994nonlinear}.
Therefore, to ensure operational safety the design of controllers for DAS and AGV applications cannot assume that these characteristics are constant or known.

Towards improving performance and robustness in case of multiplicative uncertainties (such as tire parameters), we have presented in the companion paper  \cite{massera2019tube}, the theoretical development of the tube-based guaranteed cost model predictive control (T-GCMPC) we are dealing in this paper. It can guarantee stability and feasibility robustness, and an upper bound to an MPC optimization problem cost for a linear system with multiplicative parametric uncertainties.

In this paper, we apply the T-GCMPC in an autonomous vehicle in order to avoid front and rear tire saturation and to track a provided reference trajectory up to the limits of handling of the vehicle. The proposed approach incorporates a novel cone-bounded uncertainty model-based conservative relaxation of tire force nonlinearities. It results in a linear system subject to norm-bounded multiplicative uncertainties approximation of the nonlinear vehicle model.
We also propose a novel maximal robust controllable invariant set for the lateral vehicle control. It ensures the state belongs to the domain where such approximation is valid. 
Finally, we present both simulations and in-vehicle tests of the proposed approach. A comparative study between the proposed invariant set and the set used by \cite{beal2013mpc} and \cite{funke2017collision} is also presented.

The remainder of the paper is organized as follows: in Section \ref{sec_vehicle_model}, we present the vehicle modeling and discuss the tire models and its uncertainties; in Section \ref{sec_invariant_set}, we formulate the proposed robust controllable invariant set and we compare it to previously proposed ones; in Section \ref{sec_controller}, we describe the proposed tube-based guaranteed cost model predictive controller formulation; in Section \ref{sec_experiments}, we report the simulations and experimental tests performed; and in Section \ref{sec_conclusion}, we provide final remarks.

\section{TIRE AND VEHICLE MODELS}
\label{sec_vehicle_model}

The vehicle model dealt with in this paper  incorporates elements of the tire dynamics as uncertainties. The brush tire model presented in \cite{fiala1987kraftfahrzeugtechnik} assumes a parabolic pressure distribution on the contact patch whose rubber dynamics has reached steady-state. Under such conditions, the lateral force for a given tire $ i $ becomes a direct function of the slip angle ($\alpha_i$). It also depends on  the cornering stiffness ($ C_{i} $), the static friction coefficient ($ \mu_i $), the tire normal force ($ F_{z_i} $) and the ratio between dynamical and static friction coefficients ($ R_{\mu_i} $). Then, the lateral force is given by%
\begin{equation}
F_{y_i} = \left\{\begin{matrix}
a_i f_i + b_i |f_i| f_i + c_i f^3_{i} & | \alpha_i | \le \alpha_i^{sat}\\ 
F_{sliding,i} \text{sgn}(\alpha_i)& otherwise
\end{matrix}\right. 
\label{eq_fiala_tire}
\end{equation}%
where%
\begin{align}
f_i &= tan(\alpha_i),\;\;\;\;a_i = - C_i, \nonumber \\
b_i &= k_{\mu_i} C_i^2 \frac{2 - R_{\mu_i}}{3 \mu_i F_{z_i}},\;\;\;c_i = - k_{\mu_i}^2 C_i^3 \frac{1 - \frac{2}{3} R_{\mu_i}}{(3 \mu_i F_{z_i})^{2}}, \nonumber   \\
k_{\mu_i} &= q_i - \left(\frac{2 - R_{\mu_i}}{3} - \frac{1}{9}\right) q^{2}_i,\;\;\;q_i = \left( 1 - \frac{2}{3}R_{\mu_i} \right)^{-1}, \nonumber \\
\alpha_i^{sat} &= tan^{-1}\frac{3 \mu_i F_{z_i}}{k_{\mu_i} C_{i}},\;\;\;F_{sliding,i} = - \mu_i R_{\mu_i} F_{z_i}. 
\end{align}
In this paper, we are interested in the limits of handling performance obtained at peak forces. From \eqref{eq_fiala_tire}, the peak lateral force and its respective slip angle are given by%
\begin{align}
F_{y_i}^{peak} &= \mu_i F_{z_i}\\
\alpha_{i}^{peak} &= tan^{-1}\left( \frac{q_i \mu_i F_{z_i}}{k_{\mu_i} C_{i}} \right).
\label{eq_peak_slip}
\end{align}


A linear approximation of the lateral tire forces is valid for low slip angle situations and enables the development of a linear model of the vehicle dynamics. From \eqref{eq_fiala_tire}, a first-order Taylor approximation at the point $ \alpha_i = 0 $ yields%
\begin{equation}
F_{y_i} \approx - C_i \alpha_i.
\end{equation}%
This result has been widely used by both industry and academia for decades to design controllers and estimators for DAS and AGV applications.

\subsection{Uncertain Vehicle Model}

The vehicle linear model subject to uncertainties we consider in this paper is based on the  following nonlinear bicycle model, represented by%
\begin{align}
\dot{v}_x &= \frac{1}{m} \left( F_{xf} \cos(\delta) - F_{yf} \sin(\delta) \right) + r v_y
\label{eq_nonlinear_bicycle_vxdot}\\
\dot{v}_y &= \frac{1}{m} \left( F_{xf} \sin(\delta) + F_{yf} \cos(\delta) + F_{yr}  \right) - r v_x
\label{eq_nonlinear_bicycle_vydot}\\
\dot{r} &= \frac{1}{I_z} \left( a F_{xf} \sin(\delta) + a F_{yf} \cos(\delta) - b F_{yr} \right )
\label{eq_nonlinear_bicycle_rdot}
\end{align}%
where the state is give by $ x = [v_x, v_y, r]^T$ and the control input by $ u = [F_{xf}, \delta]^T$ as$, v_x $ is the longitudinal velocity, $ v_y $ is the lateral velocity, $ r $ is the yaw rate, $ \delta $ is the front wheel steering angle, $ a $ is the distance from the center of mass to the front axle, $ b $ is the distance from the center of mass to the rear axle, $ m $ is the mass, $ I_z $ is the yaw moment of inertia, and $ F_{yf} $ and $ F_{yr} $ are the lateral tire forces described by the ``brush" tire model with respective parameters $ C_f $, $ C_r $, $ R_{\mu f} $, $ R_{\mu r} $, $ \mu =  \mu_f = \mu_r $ and normal loads $ F_{zf} = m g b / (a + b ) $ and $ F_{zr} = m g a / (a + b) $.

The tire slip angles required by the tire model are defined by the angle between the velocity vector of the contact patch and the tire direction. Therefore,%
\begin{align}
\alpha_f &= \arctan \left( \frac{v_y + a r}{v_x} \right) - \delta
\label{eq_front_slip}\\
\alpha_r &= \arctan \left( \frac{v_y - b r}{v_x} \right).
\label{eq_rear_slip}
\end{align}

The linear bicycle model is a first order Taylor approximation of its nonlinear counterpart described in \eqref{eq_nonlinear_bicycle_vxdot}-\eqref{eq_nonlinear_bicycle_rdot} at $ v_y = 0 $ and $ r = 0 $ under the assumptions $ \dot{v}_x \approx 0 $ and $ F_{xf} \approx 0 $.

We consider the tire parameter variation as a conic set uncertainty on the force domain. Then, we extend the linear bicycle model with such conic set to obtain a linear system with norm-bounded structured multiplicative uncertainties.

\begin{figure}[!ht]
\centering
\includegraphics[width=\columnwidth]{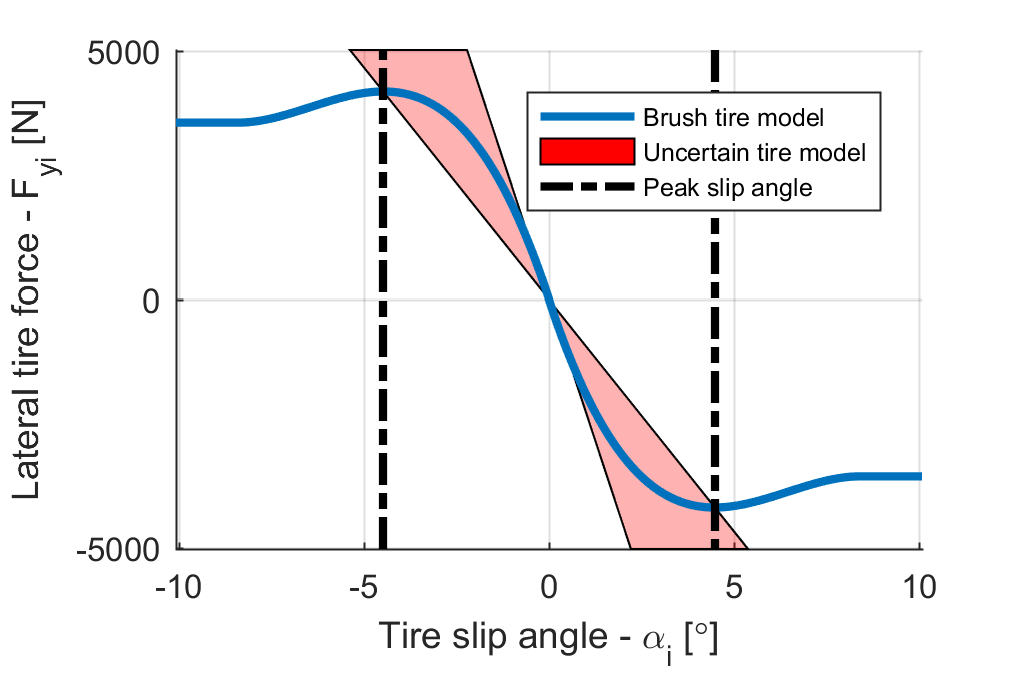}
\caption{Representation of the proposed conic-bounded uncertain tire model (red solid region) and its relation to the ``brush" tire model (blue solid line) and the slip angle with peak tire force (black dash-dot line).}
\label{fig_uncertain_tire_model}
\end{figure}

In the scope of this paper, we are interested in ensuring that both front and rear tire slip angles remain within the peaks. Therefore,%
\begin{align}
- \alpha_f^{peak} \le \alpha_f &\le \alpha_f^{peak} 
\label{eq_slip_domain_f} \\ 
- \alpha_r^{peak} \le \alpha_r &\le \alpha_r^{peak}
\label{eq_slip_domain_r}
\end{align}%
where the front, rear and and peak slip angles are defined by \eqref{eq_front_slip}, \eqref{eq_rear_slip} and \eqref{eq_peak_slip}, respectively.

The ``brush" tire model has a concave derivative within the domain of interest. Therefore, the absolute lateral tire force is upper-bounded by the tire cornering stiffness $ C_i $ and lower-bounded by the instantaneous slope defined at the peak tire force $ C_i^{peak}$, described by%
\begin{equation}
C_i^{peak} = \mu_i F_{z_i} / \alpha_i^{peak} \approx \frac{k_{\mu_i} C_i}{q_i}.
\label{eq_ci_peak}
\end{equation}%
It is worth noting that \eqref{eq_ci_peak} is approximately independent of friction based on $ tan^{-1}(x) \approx x $ used in the second term. Based on \eqref{eq_ci_peak}, we conclude that
\begin{multline}
\forall || \alpha_i || \le \alpha_i^{peak} \: \exists || \gamma_i || \le 1: \\ 
F_{yi}(\alpha_i) = - \left(\bar{C}_i + \gamma_i \delta C_i \right) \alpha_i
\label{eq_conic_representation}
\end{multline}%
where%
\begin{align}
\bar{C}_i &= \frac{C_i + C_i^{peak}}{2},\;\;\;\delta C_i = \frac{C_i - C_i^{peak}}{2}. \nonumber
\end{align}

The relaxation from \eqref{eq_conic_representation} is the proposed uncertain tire model, where $ \gamma_i $ is the norm-bounded uncertain gain related to the tire $ i $. An example of this model is shown in Figure \ref{fig_uncertain_tire_model}.


The proposed uncertain linear bicycle model is a direct application of the uncertain tire model from \eqref{eq_conic_representation} to the bicycle model \eqref{eq_nonlinear_bicycle_vxdot}-\eqref{eq_nonlinear_bicycle_rdot}, which is linearized by a first-order Taylor expansion at $ v_y = 0 $ and $ r = 0 $ under the assumptions $ \dot{v}_x \approx 0 $ and $ F_{xf} \approx 0 $. We then extend its state space to contain the vehicle position and heading in Frenet coordinates \cite{funke2016collision} for tracking a path defined in function of its curvature $ \kappa_r $. Such a system is also represented in the continuous-time linear system subject to norm-bounded uncertainties with an additional reference input:%
\begin{multline}
\dot{x}(t) = (A + B^w \Delta_k C_y) x(t) + \\ + (B^u + B^w \Delta_k D_y^u) u(t) + B^r r(t)    
\label{eq_tbm_state_space}
\end{multline}%
with matrices:%
\begin{align}
A &= \begin{bmatrix}
0 & v_x & 1 & d_m \\
0 & 0 & 0 & 1 \\
0 & 0 &  - \frac{\bar{C}_f + \bar{C}_r}{m v_x} & - \frac{a \bar{C}_f - b \bar{C}_r}{m v_x} - v_x \\ 
0&0&- \frac{a \bar{C}_f - b \bar{C}_r}{I_z v_x} & - \frac{a^2 \bar{C}_f + b^2 \bar{C}_r}{I_z v_x}
\end{bmatrix} \nonumber 
\end{align}%
\begin{align}
B^u &= \begin{bmatrix}
0 \\
0 \\
\frac{\bar{C}_f}{m} \vspace{0.2cm}\\ 
\frac{a \bar{C}_f}{I_z}\\
\end{bmatrix}, 
B^w = \begin{bmatrix}
0 & 0 \\
0 & 0 \\
-\frac{1}{m v_x} & \frac{1}{m v_x} \\ 
-\frac{a}{I_z v_x} & -\frac{b}{I_z v_x} \\
\end{bmatrix}  \nonumber
\end{align}%

\begin{align}
B^r &= \begin{bmatrix}
0 & - v_x & 0 & 0
\end{bmatrix}^T,\;\;\;
C_y = \begin{bmatrix}
0 & 0 &\delta C_f & a.\delta C_f \\ 
0 & 0 & -\delta C_r & b.\delta C_r
\end{bmatrix} \nonumber
\end{align}%
\begin{align}
D_{y}^u &=
\begin{bmatrix}
-\delta C_f\\ 
0
\end{bmatrix},\;\;\; \Delta_{k} = \begin{bmatrix}
\gamma_f & 0\\ 
0 & \gamma_r
\end{bmatrix}
\label{eq_tbm_delta}
\end{align}%
where $ x(t) = [e_y(t), e_\psi(t), v_y, r]^T $, $ u(t) = \delta(t) $, $ r(t) = \kappa_r(t) $, $ e_y(t) $ is the cross-track error of the vehicle reference point and its path, $ e_\psi(t) $ is the heading error of the vehicle body to the path tangent, and $ d_m $ is the distance along the vehicle body between the center of mass of the vehicle and the reference point where the cross-track error is measured. $ || \Delta_{k} || \le 1 $  has a known diagonal structure which the T-GCMPC synthesis will exploit to create a less conservative controller. The vertexes of the disturbance $\Delta_{k}$ can be enumerated due to its diagonal structure. This fact is in Section \ref{sec_invariant_set} to generate the maximal RCI set for the system. Such enumeration is done by evaluating the matrices $ A $ and $ B_u $ for the four permutations of the values $ \gamma_f, \gamma_r \in \{-1, 1\} $.

The uncertain system from \eqref{eq_tbm_state_space}-\eqref{eq_tbm_delta} enables the design of trajectory tracking controllers capable of operating up to the limits of the vehicle's tire forces. Therefore, the design of the proposed T-GCMPC in Section \ref{sec_controller} is developed based on this formulation.

\section{MAXIMAL ROBUST CONTROLLABLE INVARIANT SET}
\label{sec_invariant_set}

In this section we describe the proposed maximal RCI set and the algorithm for its synthesis. Its goal is to ensure that the system remains within the domain where the uncertain bicycle model is valid. Additionally, the use of RCI sets as terminal constraints on Robust MPC problems provides recursive feasibility guarantees.

\subsection{Maximal Robust Controllable Invariant Set Computation}

We consider the discrete-time linear difference inclusion system $ x_{k+1} = A x_k + B^u u_k $ \cite{boyd1994linear}, where%
\begin{equation}
\begin{bmatrix}
A & B
\end{bmatrix}
\in
\left\{
\left.
\sum_{i = 1}^{n_s} \alpha_i \begin{bmatrix}
A_i & B_i
\end{bmatrix}
\right|
\sum_{i = 1}^{n_s} \alpha_i = 1,
\alpha_i \ge 0
\right\}.
\end{equation}%
Also, let $ \mathbb{C} $ be the feasible set defined by %
\begin{equation}
\mathbb{C} = \left\{ \left. \begin{bmatrix}
x\\ 
u
\end{bmatrix} \right| H_x x + H_u u \le g \right\}
\end{equation}%
with state and control input space projections $ \mathbb{C}_x = \text{Proj}_x(\mathbb{C}) $ and $ \mathbb{C}_u = \text{Proj}_u(\mathbb{C}) $, respectively.

Let $ \mathbb{R}_0 = \mathbb{C}_x $ and $ (A_i, B_i) $ be the vertex matrices from the uncertain polytopic representation of the uncertain bicycle model. Then, the backward recursive approach to generate maximal RCIs \cite{kerrigan2001robust} is given by%
\begin{align}
\mathbb{S}_{k+1,i} &= [A_i B_i]^\dagger \mathbb{R}_k \\
\widehat{\mathbb{S}}_{k+1} &= \mathbb{C}_x \cap \left( \bigcap\limits_{i = 1}^{n_s} \mathbb{S}_{k+1,i} \right) \\
\mathbb{R}_{k+1} &= \text{Proj}_x(\widehat{\mathbb{S}}_{k+1}).
\end{align}%
Convergence is reached once $ \mathbb{R}_{k+1} = \mathbb{R}_{k} = \mathbb{R}_{\infty} $ for some $ k $. Moreover, even without finite-time convergence guarantees for such algorithm, it is still very effective for practical applications.

\subsection{Proposed Maximal Robust Controllable Invariant Set}

We generate a maximal RCI set based on the exact polytopic representation of the uncertain bicycle model from (\ref{eq_tbm_state_space}). Such a invariant set has two main goals: To ensure the tires slip angles are within the boundaries where the uncertain bicycle model approximation is valid (from \eqref{eq_slip_domain_f} and \eqref{eq_slip_domain_r}) and to ensure the commanded steering angle is within the mechanical limits of the system, given by the constraint%
\begin{equation}
- \delta_f^{max} \le \delta_f \le \delta_f^{max}.
\end{equation}%
Since the slip angles are nonlinear functions, we use the small angle approximation $ tan^{-1}(x) \approx x $ to linearize the constraints, which results%
\begin{equation}
H_x = \begin{bmatrix}
1 & a\\ 
1 & -b\\ 
0 & 0\\
-1 & -a\\ 
-1 & b\\ 
0 & 0
\end{bmatrix},
\; \; 
H_u = \begin{bmatrix}
- v_x\\ 
0\\ 
1\\
v_x\\ 
0\\ 
-1
\end{bmatrix},
\; \;
g = \begin{bmatrix}
v_x \alpha_f^{peak}\\ 
v_x \alpha_r^{peak}\\ 
\delta_f^{max}\\
v_x \alpha_f^{peak}\\ 
v_x \alpha_r^{peak}\\ 
\delta_f^{max}
\end{bmatrix}.
\end{equation}

\begin{figure}[!ht]
\centering
\includegraphics[scale=0.8]{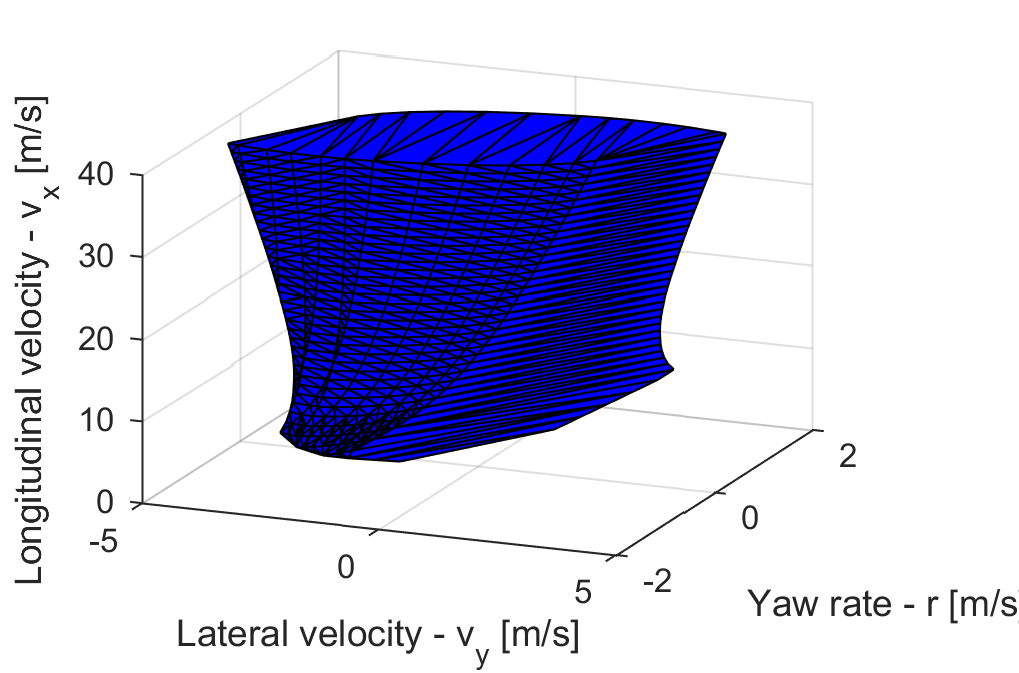}
\caption{Proposed maximal RCI set shape for on the lifted state space $ x = [v_x, v_y, r]^t $.}
\label{fig_rci_maximum}
\end{figure}

We generate the maximal RCI set for speeds from $ 3 m/s $ to $ 40 m/s $ and schedule it based on the current vehicle speed. Figure \ref{fig_rci_maximum} shows the resulting set on the space $ x = [v_x, v_y, r]^T $. We can observe that at slow speeds the RCI is constrained by the actuator limits, while at higher speeds it is bounded by slip limits in the tires. It is elongated by the slower system dynamics.

\begin{figure}[!ht]
\centering
\includegraphics[scale=0.8]{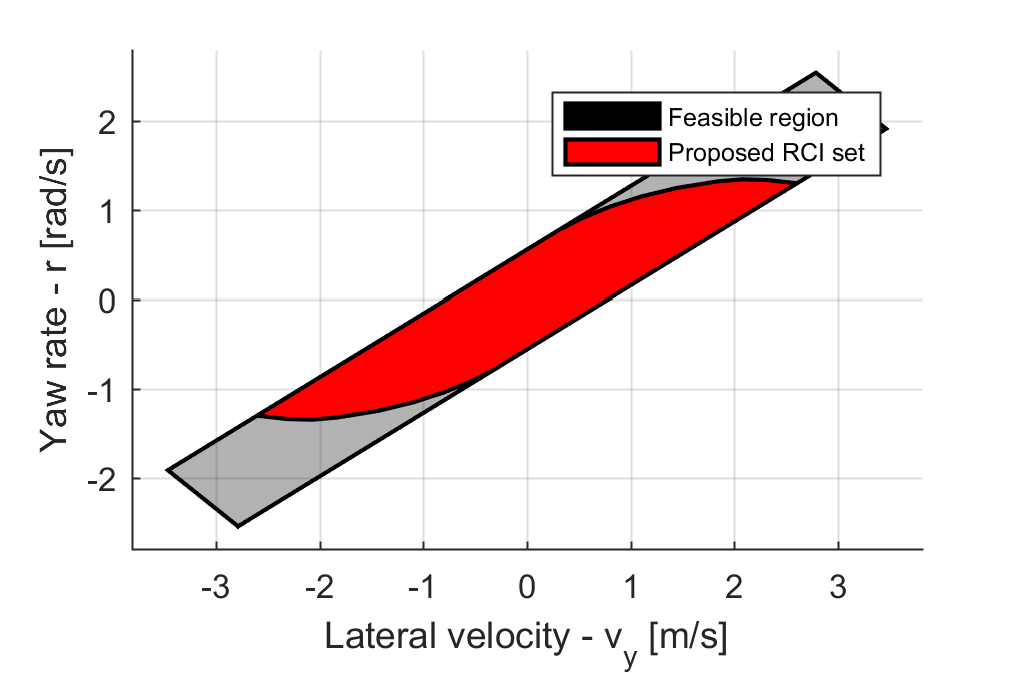}
\caption{Proposed maximal RCI set (red) at $ v_x = 10 m/s $ with its respective feasible region (grey).}
\label{fig_rci_example}
\end{figure}

Additionally, Figure \ref{fig_rci_example} shows the generated maximal RPI set for the speed $ v_x = 10 m/s $. It is possible to observe the rear peak slip angles at the boundary of the RCI, while the peak front slip angle is far from the RCI boundary. The latter is a consequence of the projection of the control input.  The steering command does not need to reach its maximum admissible value outside of sliding conditions at such speeds.

\subsection{Maximal RCI comparison}
\begin{figure}[ht]
\centering

\includegraphics[scale=0.55]{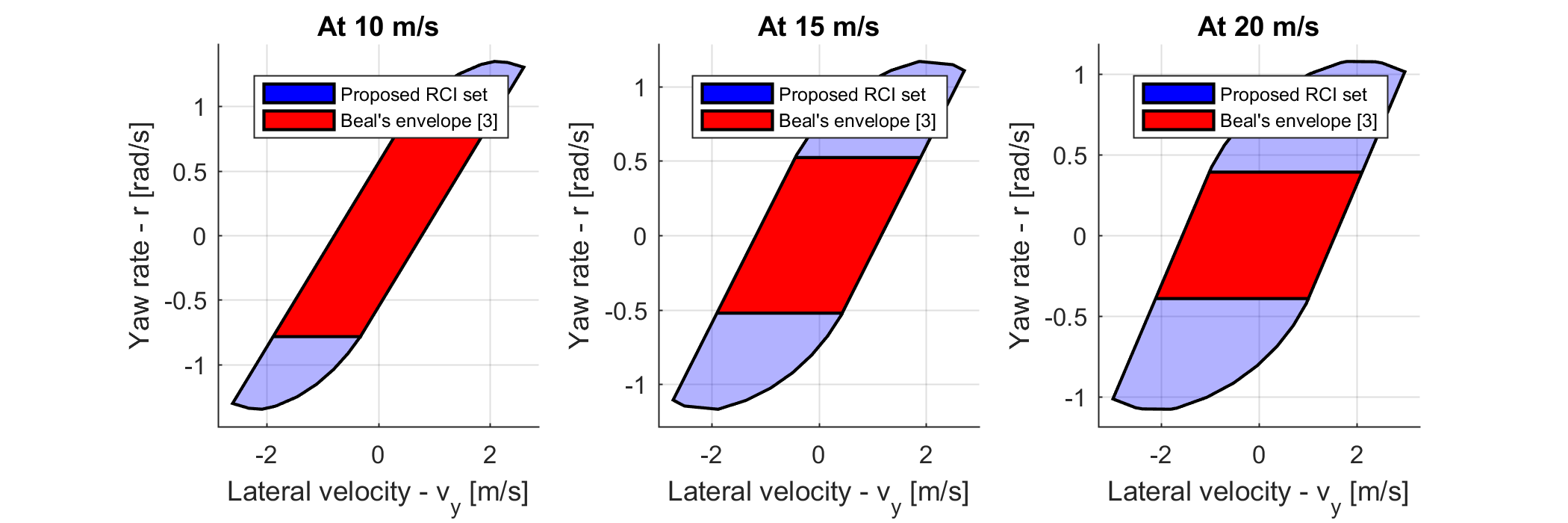}

\caption{Comparison for the proposed maximal RCI set (in blue) and Beal's envelope from \cite{beal2013mpc} (in red), at speeds $ 10 m/s $, $ 15 m/s $, and $ 20 m/s $.}
\label{fig_rci_comparison}
\end{figure}

Beal and Gerdes \cite{beal2013mpc} proposed an RCI set based on the maximum rear slip angle and the maximum achievable steady-state yaw rates, given by%
\begin{align}
- \alpha_r^{peak} \le \alpha_r \le \alpha_r^{peak} \\
- r^{max} \le r \le r^{max}
\end{align}%
where%
\begin{equation}
r^{max} = \frac{\mu}{v_{x}}\:\frac{ab + max(a,b)^{2}}{ min(a,b) (a + b)}.
\end{equation}

Figure \ref{fig_rci_comparison} shows the proposed maximal RCI set and the set proposed by Beal and Gerdes. We can observe that the peak side slip angle of the rear tire limits both sets. However, the proposed set achieves larger yaw rate values. The Beal's envelop does not achieve steady-state $ (x, u) $ pairs. Nonetheless, the proposed maximal RCI set enables the execution of highly dynamic maneuvers which momentarily exceed stable yaw rates. The model used to generate maximal RCI set is independent of the friction estimation. It is an advantage if compared with the approach proposed in \cite{beal2013mpc}.


\section{CONTROLLER FORMULATION}
\label{sec_controller}

In this section, we describe the proposed controller formulation. First, we present an overview of the T-GCMPC method proposed by the authors in \cite{massera2019tube}. Next, we describe its application for the trajectory tracking problem.

\subsection{Tube-based Guaranteed Cost Model Predictive Control}

The Tube-based Guaranteed Cost Model Predictive Control goal is to provide robust feasbility, stability, and optimality for constrained linear systems subject to multiplicative structured norm-bounded uncertainties\footnotemark[1]. Consider the uncertain discrete-time system given by%
\footnotetext[1]{We refer the reader to the GCMPC toolbox at \textit{\url{https://gitlab.com/cmasseraf/gcmpc}} for implementation details.}%
\begin{equation}
\begin{matrix*}[l]
x_{k+1} &= A_d x_k + B_d^u u_k + B_d^w w_k\\
y_k &= C_y x_k + D_y^u u_k\\
c_k &= C_c x_k + D_c^u u_k
\end{matrix*}
\label{eq_udls}
\end{equation}%
where $ x_k \in \Re^{n_x} $ is the system state, $ u_k \in \Re^{n_u} $ is the control input, $ w_k = \Delta_k y_k $ is the uncertainty input, $ y_k \in \Re^{n_y} $ is the uncertainty output, $ c_k \in \Re^{n_c} $ is the cost ouput, $ A_d  \in \Re^{n_x \times n_x} $ is the discretized system matrix, $ B_d^u \in \Re^{n_x \times n_u} $ is the discretized control input matrix, $ B_d^w \in \Re^{n_x \times n_w} $ is the discretized uncertianty input matrix, $ C_y \in \Re^{n_y \times n_x} $ is the state uncertainty matrix, $ D_y^u \in \Re^{n_y \times n_u} $ is the control input uncertainty matrix, $ C_c \in \Re^{n_c \times n_x} $ is the state cost matrix, and $ D_c^u \in \Re^{n_c \times n_u} $ is the control input cost matrix. Notice that the nomenclature of output related matrices was not changed, since exact discretization methods do not modify them.

Towards that purpose, the T-GCMPC provides a conservative solution to the intractible Min-Max MPC problem%
\begin{equation}
\begin{matrix*}[l]
J^*(x_0) = & \underset{\mathbf{u}}{\inf} \; \underset{\mathbf{\Delta}}{\sup} & \underset{N \rightarrow \infty}{\lim} J_0(x_0, \mathbf{u}, N)\\
& s.t. & x_{k+1} = A_d x_k + B_d^u u_k + B_d^w w_k\\
& & w_k = \Delta_k (C_y x_k + D_y^u u_k) \\
& & [x_k^T, u_k^T]^T \in \mathbb{C}
\end{matrix*}
\label{eq_opt_mmmpc}
\end{equation}%
where the cost function $ J_i $ is defined by%
\begin{equation}
J_i(x_i, \textbf{u}, N) = \underset{k = i}{\overset{N-1}{\sum}} x_k^T Q x_k + u_k^T R u_k + 2 x_k^T N u_k\end{equation}%
with symmetric weighting matrices $ P_N \succeq 0 $, $ Q \succeq 0 $, and $ R \succ 0 $, $ \mathbf{u} = \{u_k \mid k \in [0, N-1]\} $, and $ \mathbf{\Delta} = \{\Delta_k \mid k \in [0, N-1]\} $.

The T-GCMPC controller synthesis consists of three steps, where the first two are computed \textit{a priori} and the third is performed during the controller execution:

\subsubsection{Guaranteed Cost Controller Synthesis}
The optimal guaranteed cost controller (GCC) is a robust version of the linear quadratic regulator (LQR) \cite{xie1993control}. Its solution is a state-feedback controller $ u_k = - K x_k $ with associated symmetric matrix cost $ P \succ 0 $ with minimal trace that satisfies%
\begin{multline}
A_{cl}(\Delta)^T P A_{cl}(\Delta) - P + \\ + Q + N K + K^T N^T + K^T R K \preceq 0
\end{multline}%
for all admissible values of $ \Delta $ and where%
\begin{equation}
A_{cl}(\Delta) = A_d + B_d^w \Delta C_y - \left( B_d^u + B_d^w \Delta D_y^u \right) K.
\end{equation}

Such controller synthesis can be posed as a Semi-definite Programming (SDP) problem given by%
\begin{align}
\min \; \; & \text{trace}(Z) \\
s.t. \; \; & \\
& \begin{bmatrix}
-Z & I\\ 
 \star & -X
\end{bmatrix} \preceq 0 \\
& \begin{bmatrix}
- \Lambda_q & 0 & 0 & C_y X - D_y^u Y\\ 
\star & - I & 0 & C_c X - D_c^u Y\\ 
\star & \star & - \bar{X} & A_d X - B_d^u Y\\ 
\star & \star & \star & - X\\
\end{bmatrix} \preceq 0
\label{eq_gcc_synthesis}
\end{align}%
where $ X = P^{-1} $, $ Y = K P^{-1} $, $ \bar{X} = X - B_d^w \Lambda_p B_d^{wT} $, the cost matrices $ C_c \in \Re^{n_c \times n_x} $ and $ D_c^u \in \Re^{n_c \times n_u} $ are given by the factorization of the cost function matrices%
\begin{equation}
\begin{bmatrix}
Q & N\\ 
N^T & R
\end{bmatrix} = \begin{bmatrix}
C_c & D_c^u 
\end{bmatrix}^T \begin{bmatrix}
C_c & D_c^u 
\end{bmatrix}
\end{equation}%
or, equivalently,%
\begin{equation}
J_i(x_i, \textbf{u}, N) = \underset{k = i}{\overset{N-1}{\sum}} c_k^T c_k
\end{equation}%
and the generalized S-Procedure variables are given by%
\begin{equation}
\begin{aligned}
\Lambda_p = \text{diag}(\lambda_1 I_{n_{p1}}, \lambda_2 I_{n_{p2}}, \ldots, \lambda_s I_{n_{ps}}) \\
\Lambda_q = \text{diag}(\lambda_1 I_{n_{q1}}, \lambda_2 I_{n_{q2}}, \ldots, \lambda_s I_{n_{qs}}).
\end{aligned}
\end{equation}%
For more details on the synthesis of GCC controllers we refer to reader to Lemma 4.5 and Lemma 4.8 of \cite{massera2018optimal}.

\subsubsection{Approximate Minimal RCI Set} 

One of the main differentiating properties of robust MPCs is their approach towards the propagation of uncertainties. Tube-based methods are based on the principle of separable control policies, where the following system is considered%
\begin{equation}
\begin{matrix*}[l]
z_{k+1} &= A_d z_k + B_d^u \nu_k \\
e_{k+1} &= A_d e_k + B_d^u \rho_k + B_d^w w_k \\
y_k &= C_y (z_k + e_k) + D_y^u (\nu_k + \rho_k) \\
c_k &= C_c (z_k + e_k) + D_c^u (\nu_k + \rho_k)
\end{matrix*}
\label{eq_sys_model_separated}
\end{equation}%
where $ z_k $ defines the nominal dynamics, and $ e_k $ defines the error dynamics. Based on the identities $ x_k = z_k + e_k $ and $ u_k = \nu_k + \rho_k $, the equivalence between  \eqref{eq_sys_model_separated} and \eqref{eq_udls} is well defined.

Since $ e_k $ propagates the error, it is a unknown variable for future states. Therefore we define a RCI set $ \mathbb{E} $, associated with a controller $ K_R $ and a scaling variable $ \alpha_k \ge 0 $, such that $ e_k \in \alpha_k^2 \mathbb{E}$ , and $ \rho_k \in \alpha_k^2 (- K_R) \mathbb{E} $. This yields the following relaxed $ \alpha_k $ dynamics%
\begin{equation}
\alpha_{k+1}^2 \mathbb{E} \supseteq \alpha_k^2 (A_d - B_d^u K_R) \mathbb{E} \oplus \lambda_k^2 B_d^w \mathbb{W}
\end{equation}%
where $ \mathbb{W} = \{ w \mid w^T w \le 1 \} $, $ \lambda_k $ satisfies%
\begin{equation}
\begin{bmatrix}
z_k \oplus \alpha_k^2 \mathbb{E} \\
\nu_k \oplus \alpha_k^2 (- K_R) \mathbb{E}
\end{bmatrix}
\subseteq \lambda_k^2 \mathbb{Y} \end{equation}%
and%
\begin{equation}
\mathbb{Y} = \left\{ \left.
\begin{bmatrix}
x \\ u
\end{bmatrix}
\right|
(\bullet)^T (C_y x + D_y^u u) \le 1
\right\}.
\end{equation}

To minimize the overall conservativeness of the controller, it is proposed a minimal RCI (mRCI) set that bounds errors dynamics. However, computing such a set is often intractable. Therefore, many approximate approaches were proposed in the literature, see for instance \cite{ kouramas2005minimal} and \cite{rakovic2005invariant}. The T-GCMPC synthesis proposes a novel approximate mRCI set synthesis. It is valid for a subset of the state space, even when the  approximate set is  subject  to any arbitrary scaling. The mRCI set $ \mathbb{E} = \{ x \mid x^T E_R x \le 1 \} $ is the result of the following SDP optimization:%
\begin{align}
\Theta_a(a_\alpha) = \inf & \tr(X)
\label{eq_mrpi_synthesis} \\ 
s.t. & \begin{bmatrix}
- X & A_d X - B_d^u Y & B_d^w \Upsilon_p\\ 
\star & -X a_\alpha & 0\\ 
\star & \star & - \Upsilon_p
\end{bmatrix} \preceq 0
\nonumber \\
& \begin{bmatrix}
- I_{n_{qi}} & C_{y,i} X - D_{y,i}^u Y\\ 
\star & - X
\end{bmatrix} \preceq 0
\nonumber \\
& \begin{bmatrix}
- \upsilon_i & 1\\ 
\star & - a_{\sigma_i}
\end{bmatrix} \preceq 0 
\nonumber \\
& a_\alpha + \sum\limits_{i = 1}^{s} a_{\sigma_i} \le 1
\nonumber
\end{align}%
where $ E_R = X^{-1} $ and $ K_R = Y X^{-1} $ and the generalized S-Procedure variables are given by%
\begin{align}
\Upsilon_p &= \text{diag}(\upsilon_1 I_{n_{p1}}, \upsilon_2 I_{n_{p2}}, \ldots, \upsilon_s I_{n_{ps}})\\
\Upsilon_q &= \text{diag}(\upsilon_1 I_{n_{q1}}, \upsilon_2 I_{n_{q2}}, \ldots, \upsilon_s I_{n_{qs}}).
\end{align}

\subsubsection{T-GCMPC Optimization}

Based on the results of the GCC synthesis \eqref{eq_gcc_synthesis}, the approxiamate mRPI synthesis \eqref{eq_mrpi_synthesis}, and the terminal RCI set $ \mathbb{R}_N = \{ x \mid H_N x \le g^N \} $ (to guarantee recursive feasibility), we can pose the T-GCMPC optimization problem as%
\begin{equation}
\begin{matrix*}[l]
& \underset{\mathbf{\nu}}{\inf} & x_0^T P x_0 + \sum\limits_{k = 0}^{N - 1} \gamma_k^2\\
& s.t. & z_{k+1} = (A_d - B_d^u K) z_k + B_d^u \nu_k\\
& & \alpha_{k+1} \ge \left|\left|\left[
\sqrt{a_\alpha} \alpha_{k}, \sqrt{a_{\sigma_1}} (\sigma_k)_1, \dots, \sqrt{a_{\sigma_s}} (\sigma_k)_s \right]\right|\right|_2 \\ 
& & (\sigma_k)_i \ge || (\bar{C}_y)_i z_k + (D_y^u)_i \nu_k || + (C_y^\alpha)_i \alpha_k\\
& & \gamma_k \ge || \bar{R}^\frac{1}{2} \nu_k ||_2 + || \bar{R}^\frac{1}{2} (K_R - K) E_R^{-\frac{1}{2}} ||_2 \alpha_k \\
& & \bar{H}_i z_k + (H_u)_i \nu_k + || (\bar{H}_R)_i E_R^{-\frac{1}{2}} ||_2 \alpha_k \le g_i \\
& & H^N_i z_N  + || H^N_i E_R^{-\frac{1}{2}} ||_2 \alpha_N \le g^R_i
\end{matrix*}
\label{eq_opt_tgcmpc}
\end{equation}%
where $ \bar{H} = H_x - H^u K $, $ \bar{H}_R = H_x - H^u K_R $, and the resulting control command $ u_0 = - K x_0 + \nu_0 $.

The T-GCMPC optimization \eqref{eq_opt_tgcmpc} is a conservative approximation to the min-max MPC problem \eqref{eq_opt_mmmpc}. It results in a second order coning programming  optimization problem; nonetheless, the increase in complexity caused by the robustness formulation grows linearly with the number of uncertainties. This is one of the main advantages of tube-based approaches. Meanwhile, other RMPC methods may grow exponentially \cite{rakovic2013homothetic, scokaert1998min} or quadratically  \cite{massera2017guaranteed, rakovic2012parameterized}.

For a detailed explanation and proof of \eqref{eq_opt_tgcmpc} and on the synthesis of the approximate mRPI set in \eqref{eq_mrpi_synthesis}, we refer the reader to \cite{massera2019tube}.

\subsection{Functional Modelling and Controller Synthesis}

One of the fundamental aspects of optimal control design is the  choice of a cost function and weighting terms. It directly impacts its nominal region performance where constraints are inactive. Towards the purpose, we choose to apply the implicit model following (IMF) \cite{tyler1964characteristics} approach to the cost function definition.

Consider the trajectory-relative bicycle model without uncertainties. Then, the goal of the controller is to minimize the cross-track error with a damped response. Therefore, we choose a first order model with time-constant $ \tau $, such that%
\begin{equation}
\dot{e}_y^r(t) = - \tau^{-1} e_y^r(t)
\end{equation}%
which, given $ \bar{C}_c = [1\; 0\; 0\; 0] $ and $ C_c B^u = 0$, is equivalent to%
\begin{multline}
\bar{C}_c (A + \tau^{-1} I) x(t) + \bar{C}_c B^u u(t) = \\ = \bar{C}_c (A + \tau^{-1} I) x(t) = 0.
\label{eq_imf_manifold}
\end{multline}%
However, \eqref{eq_imf_manifold} only defines a manifold of the system state space. Therefore, we introduce the error variable $ c(t) = C_c x(t) + D_c^u u(t) $ such that%
\begin{equation}
C_c = \begin{bmatrix}
\bar{C}_c (A + \tau^{-1} I) \\ 0
\end{bmatrix},
\; \; \;
D_c^u = \begin{bmatrix}
0 \\ 1
\end{bmatrix}
\end{equation}%
where the first row of $ c(t) $ defines the deviation from the IMF manifold from \eqref{eq_imf_manifold} while the second defines the control input ``effort".

\begin{figure}[ht]
\centering
\includegraphics[scale=0.55]{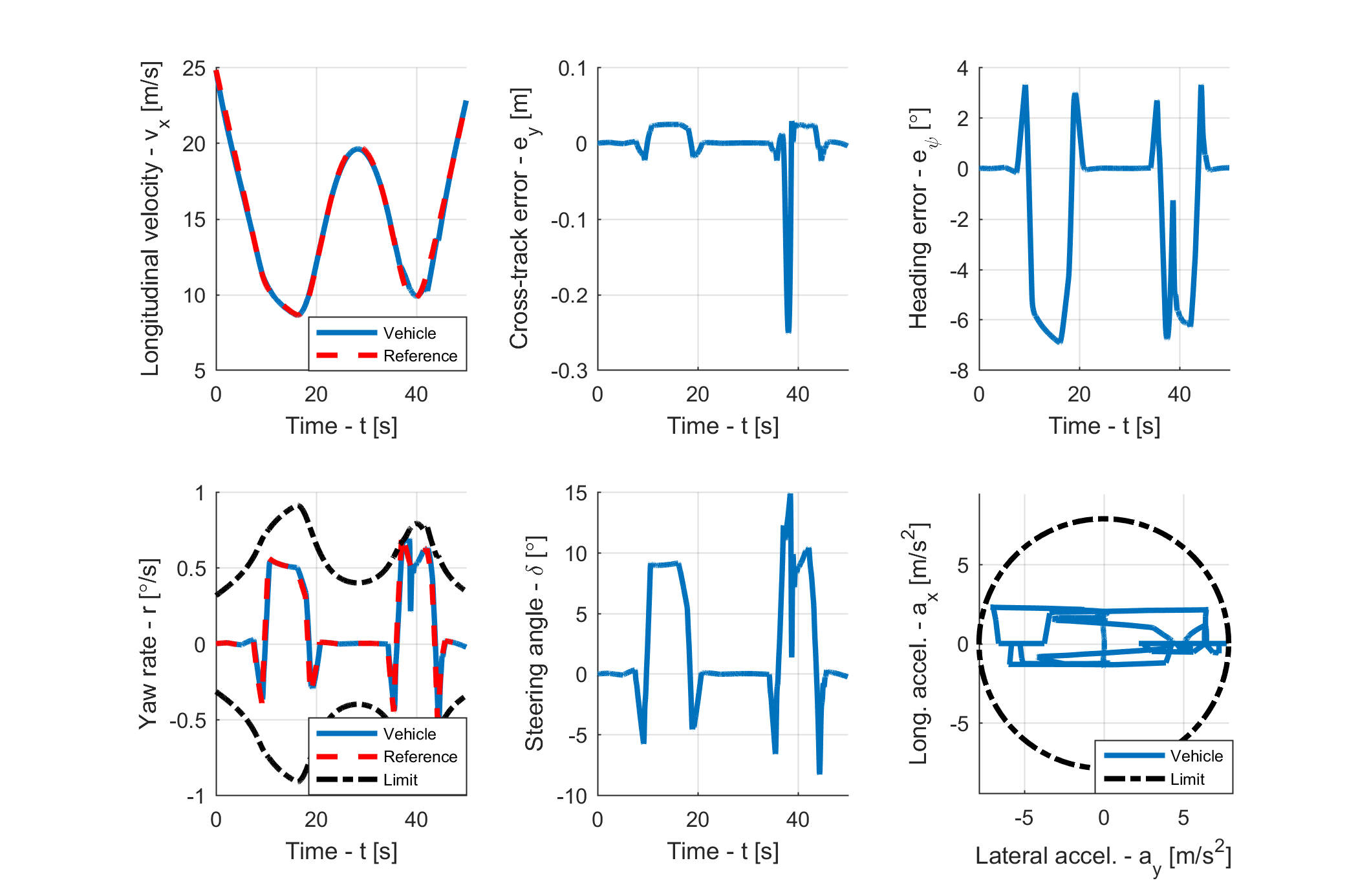}
\caption{Simulation results. The left top graph shows the longitudinal speed $ v_x $ in $ m/s $. The center top graph shows the crosstrack error $ e_y $ in $ m $. The right top graph shows the heading error $ e_\psi $ in $ \circ $. The left bottom graph shows the vehicle yaw rate (blue solid line), the reference yaw rate (dashed red line) and the yaw rate steady-state limits (dash-dot block line) in $ \circ/s $. The center bottom graph shows to commanded steering angle $ \delta_f $ in $ \circ $. The right bottom graph shows g-g diagram with longitudinal and lateral acceleration $ m/s^2 $.}
\label{fig_sim_results}
\end{figure}

\begin{figure}[!ht]
\includegraphics[scale=0.55]{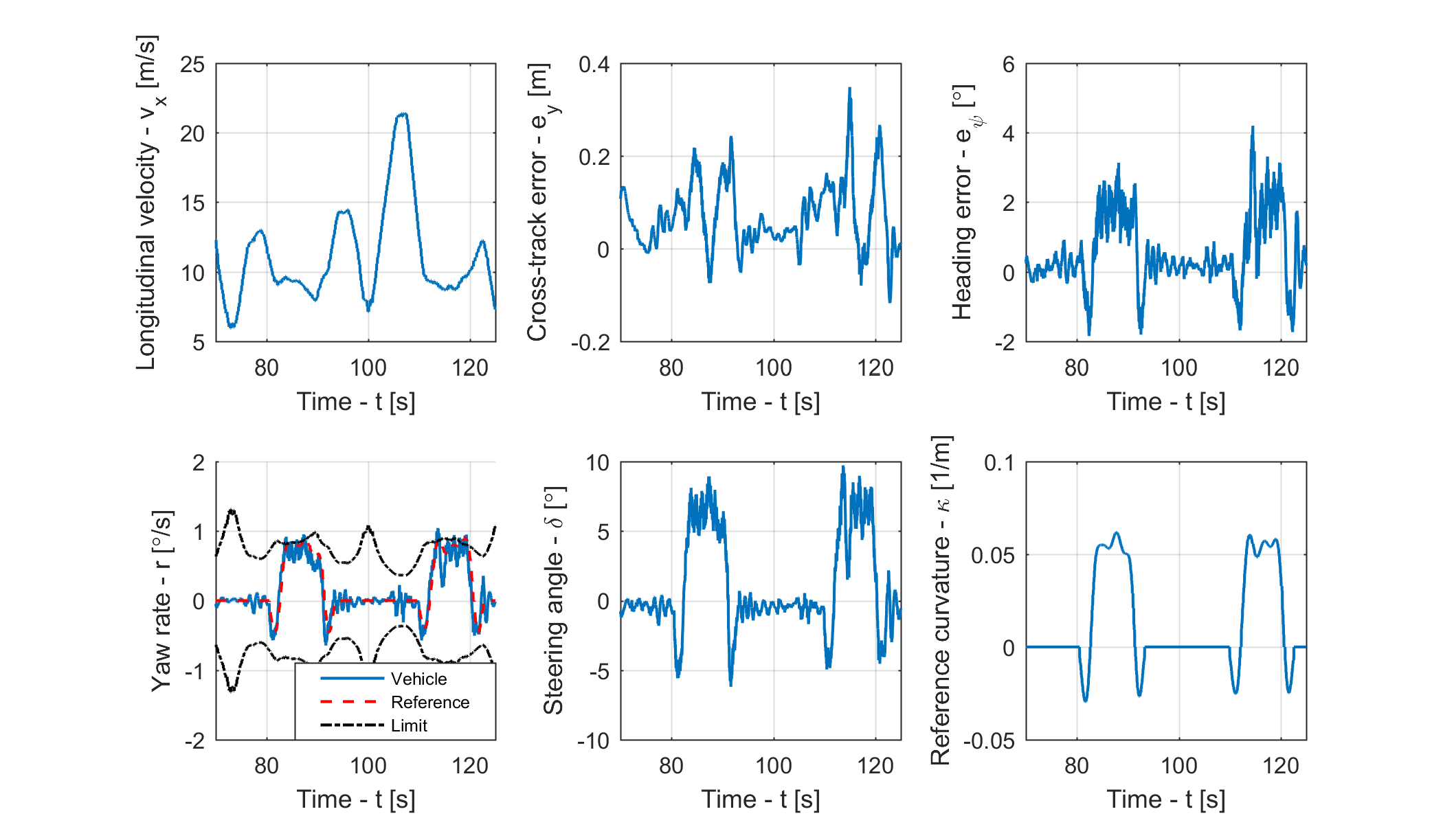}
\caption{Actual results. The left top graph shows the longitudinal speed $ v_x $ in $ m/s $. The center top graph shows the crosstrack error $ e_y $ in $ m $. The right top graph  shows the heading error $ e_\psi $ in $ \circ $. The left bottom graph shows the vehicle yaw rate (blue solid line), the reference yaw rate (dashed red line) and the yaw rate steady-state limits (dash-dot block line) in $ \circ/s $. The center bottom graph shows to commanded steering angle $ \delta_f $ in $ \circ $. The right bottom graph shows the path reference curvature $ \kappa $ in $ 1/m $.}
\label{fig_vehicle_results}
\end{figure}

The T-GCMPC synthesis requires discrete-time dynamics; meanwhile, the uncertain bicycle model is defined in continuous time. Therefore, we perform an exact discretization method, which results in a system in the form \eqref{eq_udls}.

As mentioned in Section \ref{sec_invariant_set}, the controller is synthesized for several longitudinal velocities, as the bicycle model is parametrized by $ v_x $. For each synthesis a controller gain $ K $ is obtained based on the GCC \eqref{eq_gcc_synthesis}. An approximated mRPI set is calculated from \eqref{eq_mrpi_synthesis} and an optimization problem of the form \eqref{eq_opt_tgcmpc} is posed with $ x_0 = z_0$ and $ \alpha_0 = 0$. Additionally, all constraints are modeled as soft due to possible effects of unmodeled disturbances.

Finally, Table \ref{tab_vehicle_parameters}  shows the vehicle and controller parameters used for synthesis.

\begin{table}[thpb]
    \centering
    \caption{Vehicle and controller parameters}
    \label{tab_vehicle_parameters}
    \renewcommand{\arraystretch}{1.2}
    \begin{tabular}{|lccc|}    
    \hline
    Parameter  & Symbol  & Value & Unit \\
    \hline
    Vehicle mass & $ m $ & 1231 & $ Kg $ \\
    Inertia moment $ z $ & $ I_{z} $ & 2034.5 & $ \frac{Kg}{m^{2}} $ \\
    Front axle distance to CG & $ a $ & 1.07 & $ m $ \\
    Back axle distance to CG & $ b $ & 1.40 & $ m $ \\
    Front tire cornering stiffness & $ C_{f} $ & 100000 & $ \frac{N}{rad} $ \\
    Rear tire cornering stiffness & $ C_{r} $ & 130000 & $ \frac{N}{rad} $ \\
    Peak front tire cornering stiffness & $ C_{f}^{peak} $ & 41171 & $ \frac{N}{rad} $ \\
    Peak rear tire cornering stiffness & $ C_{r}^{peak} $ & 53522 & $ \frac{N}{rad} $ \\
    Tire-road friction coefficient  & $ \mu $ & 0.8 & - \\
    Friction coefficient ratio & $R_{\mu,i}$ & 0.85 & - \\
    Prediction horizon & $ N $ & $ 10 $ & $ - $ \\
    IMF time constraint & $ \tau $ & 1 & $ s $ \\
    IMF manifold weight & $ W_{imf} $ & $ 0.087 $ & $ s^{2} / m^{2} $ \\
    Steering angle weight & $ W_{\delta_f} $ & $ 1 $ & $ 1 / rad^{2} $ \\
    Maximum front slip angle & $ \alpha_f^{peak} $ & $ 7.5760 $ & $ \circ $ \\
    Maximum rear slip angle & $ \alpha_r^{peak} $ & $ 4.4711 $ & $ \circ $ \\
    Sampling Time & $T_s$ & $0.025$ & s\\
    \hline
    \end{tabular}
\end{table}

\section{EXPERIMENTAL RESULTS}
\label{sec_experiments}

\begin{figure}[!t]
\centering
\includegraphics[width=5cm]{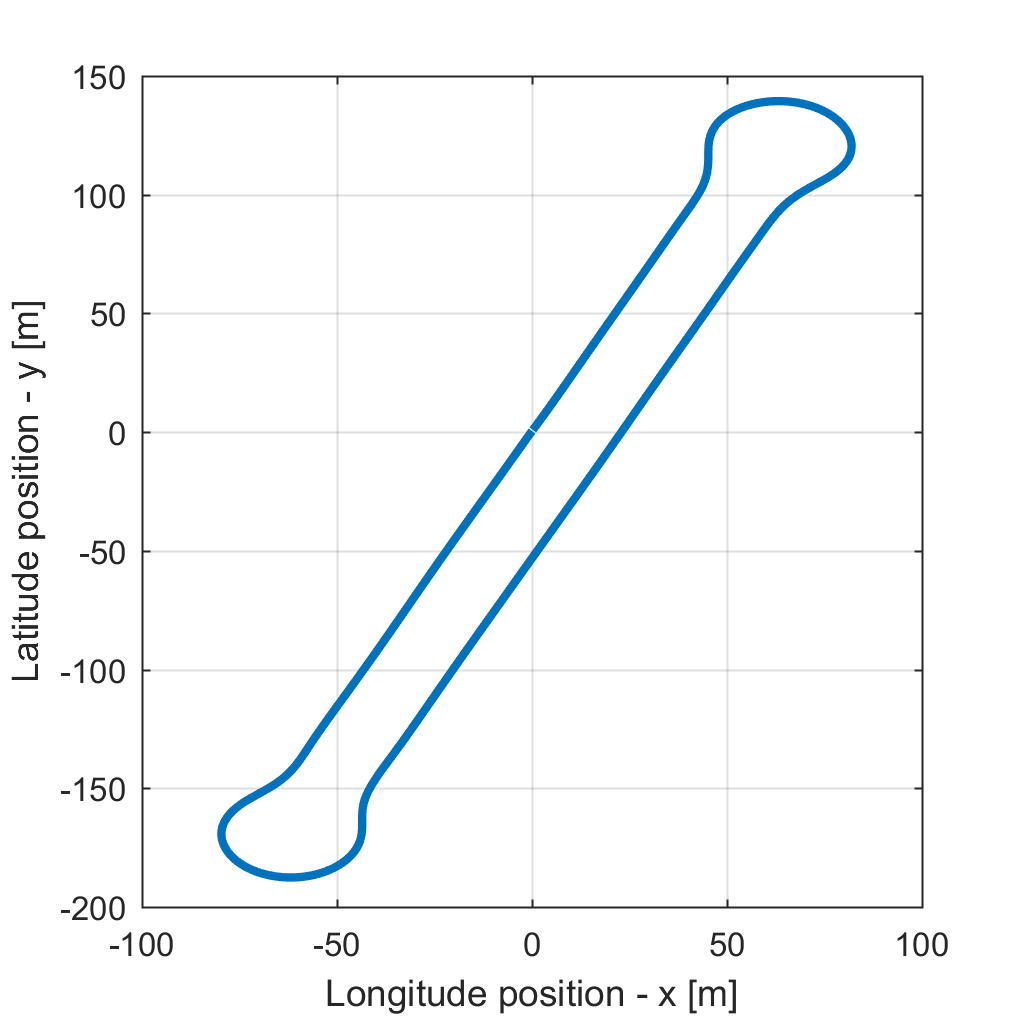}
\caption{Route used for in-vehicle experiments.}
\label{fig_route}
\end{figure}

In this section, we present results for in-vehicle test experiments. We consider a $ 750 m $ route consisting of two left turns, where the vehicle must turn right to exit and enter such turn, as shown in Figure \ref{fig_route}.

\subsection{Simulation Experiments}

Figure \ref{fig_sim_results} shows simulation results. The route is equivalent to the in-vehicle experiments; however, the simulation results have proportional-integral (PI) controller with feed-forward acceleration term implemented to control the longitudinal speed of the vehicle. The vehicle saturates the front tire on the second turn, where the highest lateral acceleration is required to follow the predefined trajectory, causing the controller to quickly counter-steer to avoid an over-steering condition. At this point we also observe the peak cross-track error of $ e_y = -0.24m $. Meanwhile, when the controller is subject to other operation conditions, the cross-track error remains bounded to $ e_y \in [-0.03, 0.03] m $.

\subsection{In-Vehicle Experiments}

\begin{figure}[!ht]
\centering
\includegraphics[width=5cm]{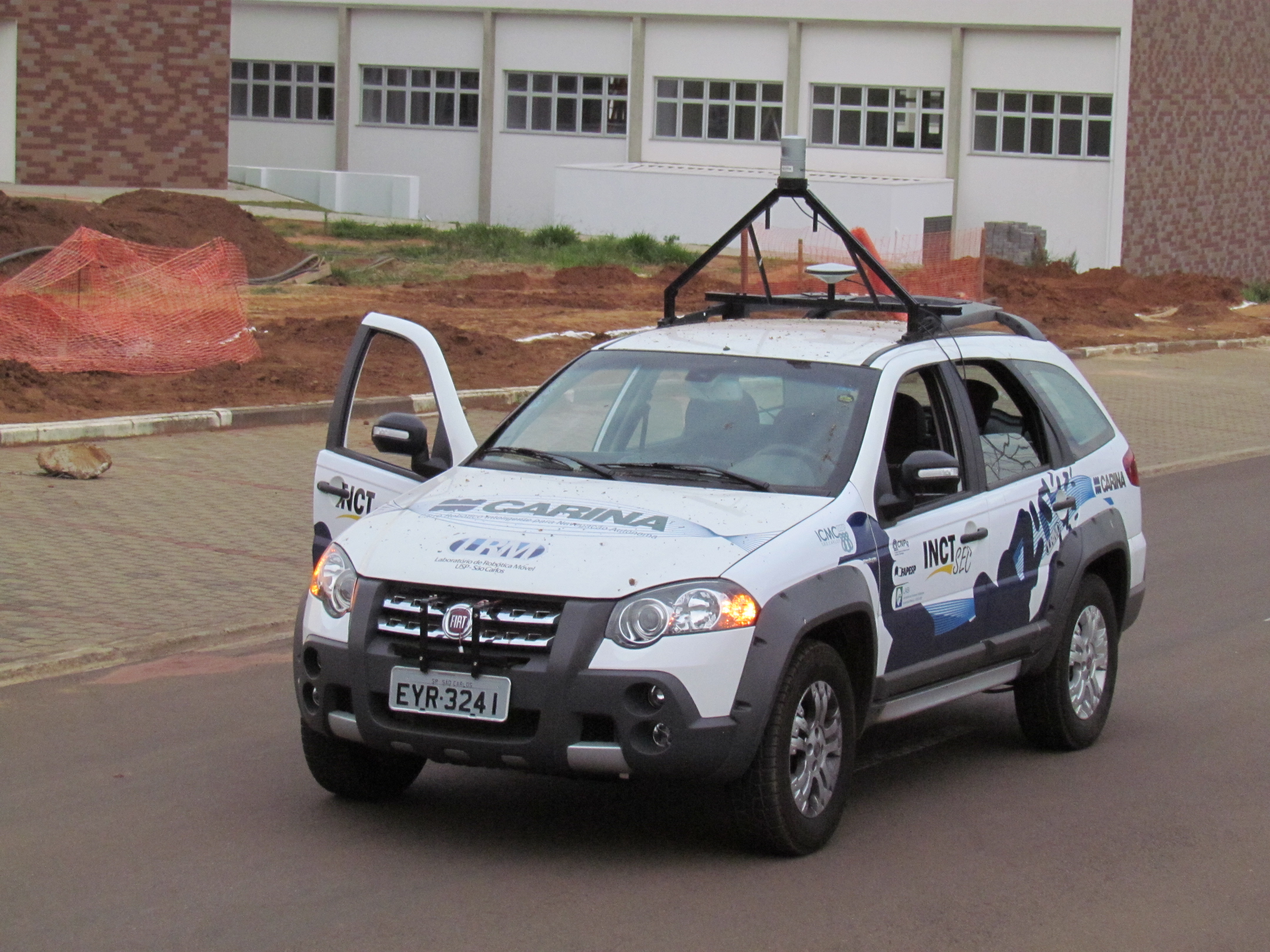}
\caption{CaRINA II platform used in experiments, a Fiat Palio Adventure.}
\label{fig_carina2}
\end{figure}

We perform the in-vehicle experiments in the CaRINA II platform \cite{shinzato2016carina}, shown in Figure \ref{fig_carina2}. This platform is a 2011 Fiat Palio Adventure modified for computational control of steering, throttle, and brake. Cameras and a 3D LIDAR sensors are installed and used by the onboard computer running Ubuntu Linux and ROS (Robot Operating System). A Septentrio GPS System provides vehicle localization and speed information with RTK correction and IMU integration operating at 10Hz, while lateral velocity and yaw rate information is estimated based on a fixed gain-scheduled observer in Frenet coordinates.

Figure \ref{fig_vehicle_results} shows the results for the in-vehicle execution. The driver manually controls the vehicle velocity while the proposed T-GCMPC controls the steering actuator. The vehicle achieves  maximum lateral acceleration of $ 8.3 m/s^2 $ around $ 113 s $ where the vehicle exceeds the maximum steady-state yaw rate. It  results in a fast steering correction to recover vehicle stability. Such a saturation also results in a peak cross-track error of $ 0.35 m $ and heading error of $ 4.2 $ degrees. The turn where such a peak occurs is an off-camber turn which is a source of unmodeled additive disturbance that contributes to such saturation.

Other high lateral acceleration conditions happen  between $ t \in [83, 90] s$. It  results in cross-track errors in the range $ e_y \in [-0.07, 0.22] m $ where the vehicle maintained itself within stability limits. Finally, during non-limits of handling conditions the vehicle maintains cross-track errors within $ e_y \in [-0.04, 0.13] m $.

Three main sources of disturbance contribute to the observed performance difference between simulation and in-vehicle are:%
\begin{enumerate}
\item State estimation are used to obtain lateral velocity and yaw rate estimates for in-vehicle tests, while simulation has noise-free full state measurement;
\item Intrinsic actuation and computational delays when operating in the vehicle; and
\item Unaccounted additive disturbances, such as steering actuation and heading misalignment, banking angle influence, and others.
\end{enumerate}%
Nonetheless, the simulation to in-vehicle performance degradation was $ \approx 0.15m $.

\section{CONCLUSIONS}
\label{sec_conclusion}

In this paper, we proposed a tube-based guaranteed cost model predictive controller for autonomous vehicles. It is able to avoid front and rear tire saturation and to track a provided reference trajectory up to the limits of handling of the vehicle. We presented a novel conservative approximation of the nonlinear vehicle dynamics to a linear system subject to norm-bounded multiplicative uncertainties;
 a maximal RCI set for vehicle dynamics, which contains a larger region of the state space when compared to other state-of-the-art invariant sets;
 an application of Tube GCMPC approach, proposed by the authors in \cite{massera2019tube}, to the trajectory-tracking problem of autonomous vehicles.
We have also presented results for both simulation and in-vehicle experiments, where we observed both low error performance and tire saturation avoidance.

Future work consists of extending the proposed controller to consider differential braking capabilities of ESC systems and integrate both longitudinal and lateral controllers.

\bibliographystyle{IEEEtranS}
\bibliography{refs/bib/paper}

\end{document}